\documentclass[final,3p,times,onecolumn]{elsarticle}
\usepackage{lineno,hyperref}

\usepackage{amsmath}
\usepackage{graphicx,epsfig,wrapfig}
\usepackage[caption=false]{subfig}

\usepackage{amsthm,mathrsfs}
\usepackage{makeidx}
\usepackage{amsfonts}
\usepackage{amssymb}
\usepackage{amsmath}
\usepackage{mathtools}
\usepackage{fixmath}
\usepackage{amsbsy}
\usepackage[capitalize]{cleveref}
\usepackage{mathrsfs}
\usepackage{slashed}
\usepackage{bm}
\usepackage{multirow}
\usepackage{booktabs}
\usepackage{adjustbox}

\usepackage{color}
\usepackage[normalem]{ulem}

\definecolor{myGreen}{rgb}{0.2,0.72,0.2}
\renewcommand\sout{\bgroup \color[rgb]{0.55,0.00,0.99} \ULdepth=-.5ex \ULset}

\newcommand{\green}[1]{\textcolor{myGreen}{#1}}

\renewcommand{\l}{\lambda}
\newcommand{\m}{\mu}
\newcommand{\lb}{{\bar{\lambda}}}
\newcommand{\mb}{{\bar{\mu}}}

\renewcommand{\L}{\mathcal{L}}

\usepackage{simplewick}

\usepackage{multirow}
\usepackage{braket}
\usepackage{soul}

\newcommand{\ta}{\left(}
\newcommand{\qa}{\left[}

\newcommand{\tc}{\right)}
\newcommand{\qc}{\right]}

\newcommand{\rDer}[1]{\overset{\rightarrow}{#1}\phantom{\,}}

\newcommand{\e}{\epsilon}

\renewcommand{\[}{\begin{equation}}
\renewcommand{\]}{\end{equation}}

\definecolor{darkgreen}{RGB}{0,120,0}

\allowdisplaybreaks

\modulolinenumbers[5]
\journal{Physics Letters B}
\biboptions{sort&compress}
\begin{document}
\begin{frontmatter}

\title{Collinear matching for next-to-leading power transverse-momentum distributions}%\tnoteref{t1}

\author[SRaddr]{Simone Rodini\corref{mycorrespondingauthor}}
\cortext[mycorrespondingauthor]{Corresponding author at: CPHT,  Ecole polytechnique, Institut Polytechnique de Paris, 91120 Palaiseau, France}
\ead{simone.rodini@polytechnique.edu}
\author[BPaddr1]{Alessio Carmelo Alvaro}
\author[BPaddr1,BPaddr2]{Barbara Pasquini}
%\ead{barbara.pasquini@unipv.it}

\address[SRaddr]{CPHT, CNRS, Ecole polytechnique, Institut Polytechnique de Paris, 91120 Palaiseau, France}
\address[BPaddr1]{Dipartimento di Fisica, Universit\`a degli Studi di Pavia, I-27100 Pavia, Italy}
\address[BPaddr2]{Istituto Nazionale di Fisica Nucleare, Sezione di Pavia, I-27100 Pavia, Italy}

\begin{abstract}
Phenomenological studies of transverse momentum dependent (TMD) parton distributions rely on the expansion in small values of the transverse separation of fields, where TMD parton distributions match onto collinear parton distribution functions. In this work, we derive this expansion at tree-level for the genuine next-to-leading power quark-gluon-quark TMDs, taking into account all the target mass corrections. We find that only a limited number of TMD parton distributions exhibit matching to twist-three collinear distributions, leading to a significant simplification in the analysis of the structure functions of semi-inclusive deep inelastic scattering.
\end{abstract}

\date{\today}


\end{frontmatter}

%%%%%%%%%%%%%%%%%%%%%%%%%%
\section{Introduction}
\label{s:introduction}
%%%%%%%%%%%%%%%%%%%%%%%%%%
Recently, significant progress has been made in advancing the next-to-leading power (NLP) factorization program for Drell-Yan and SIDIS processes in the small transverse momentum region, achieving one-loop accuracy \cite{Vladimirov:2021hdn,Rodini:2023plb}. 
In the small transverse momentum region, Drell-Yan and SIDIS processes are described in terms of transverse-momentum dependent (TMD) distributions. 
These distributions are conveniently examined in a mixed coordinate-momentum space, given by  the transverse separation of the fields $b$ (Fourier conjugated to the transverse momentum) and the (fraction of) longitudinal momentum.
In contrast to the relatively straightforward structure of the leading-power (LP) factorization theorem, the NLP approach introduces a more intricate framework. There are $32$ NLP TMD parton distributions and exactly half of them participate to SIDIS and Drell-Yan cross sections. Furthermore, NLP factorization theorem entails addressing subtle issues, such as the cancellation of special rapidity divergences that arise between different terms of the factorization theorem \cite{Rodini:2022wki}.
Notably, as the field separation $b$ approaches zero, the TMD parton distributions become the more familiar collinear parton distribution functions (PDFs). The systematic procedure that establishes the connection between these two types of distributions is known as matching, which is a form of operator product expansion (OPE) of the TMD operator. 
\\ The small-$b$ matching for LP TMD distributions is known for all distributions at least at next-to-leading order (NLO) accuracy \cite{Moos:2020wvd,Rein:2022odl}, but for some specific cases a greater precision has been achieved
\cite{Ebert:2020qef,Luo:2020epw,Gutierrez-Reyes:2018iod, Gutierrez-Reyes:2019rug,Gutierrez-Reyes:2017glx,Buffing:2017mqm, Bacchetta:2013pqa,Kanazawa:2015ajw,Scimemi:2018mmi, Meissner:2007rx,Boer:2003cm,Ji:2006ub,Zhou:2009jm,Dai:2014ala,Scimemi:2019gge}. 
The small-$b$ matching is an essential tool for practical TMD phenomenology. By constraining the behavior of the distributions at small transverse separation,  the predictive
power of the formalism increases. In fact, all modern phenomenological extractions of TMDs are based on
these relations \cite{Bacchetta:2017gcc,Scimemi:2017etj, Scimemi:2019cmh,Bacchetta:2022awv, Cammarota:2020qcw, Echevarria:2020hpy, Bury:2022czx,Cerutti:2022lmb,Moos:2023yfa}.
Our aim with this work is to establish the tree-level matching for the NLP TMD parton distributions at the leading-twist accuracy. 
%Extending the analysis beyond leading-twist accuracy would likely not have practical significance, as it would involve twist-4 PDFs that are  of limited relevance for any foreseeable phenomenological study.
Extending the analysis beyond leading-twist accuracy would probably lack practical significance. This extension involves twist-4 PDFs, which in turn involve explicitly a significant number of non-quasi-partonic operators. However, these efforts are likely to provide minimal, if any, valuable insights in terms of phenomenology.

The work is structured as follows.
In Sec.~\ref{sec_Definitions} we review the fundamental definitions of the NLP TMD distributions as well as the twist-three collinear parton distributions. In Sec.~\ref{sec_Overview_method} we discuss the basic principles of the method employed  to derive the small-$b$ matching series.
The results for the matching of NLP TMD parton distributions onto collinear twist-3 parton distributions are presented in  Sec.~\ref{sec_Results}, discussing also  their impact on the SIDIS structure functions. \green{We report an example of the computation in the appendix.}
%%%%%%%%%%%%%%%%%%%%%%%%%%
\section{Definitions}
\label{sec_Definitions}
%%%%%%%%%%%%%%%%%%%%%%%%%%
Let us introduce two light-cone vectors $n,\bar{n}$ such that $n^2 = \bar{n}^2=0$, $(n\bar{n}) \equiv n^\mu\bar{n}_\mu = 1$. The transverse space is identified by the condition $(vn)=(v\bar{n})=0$, for all $v$ vectors of the space.
TMD operators at LP and NLP can be built from semi-compact operators of definite geometric twist \cite{Vladimirov:2021hdn,Rodini:2022wki}. The `semi-compact' qualifier refers to the presence of infinite-length Wilson line in the definition of the operators. For TMD operators up to NLP, the following semi-compact operators of twist-1 and twist-2 are sufficient:
\begin{eqnarray}
\label{definition_SemiCompactOperators-1}
U_{1,i}(z;b) &=& \mathbf{W}_{\!\L}(b)[\L n +b,z n+b]q_i(z n+b), \\
U^\mu_{2,i}(z_2,z_1;b) &=& \mathbf{W}_{\!\L}(b)[\L n +b,z_2n+b] gF^{\mu+}(z_2n+b)[z_2n+b,z_1n+b]q_i(z_1n+b),\\
\overline{U}_{1,i}(z;b) &=& \bar{q}_i(z n+b)[z n+b,\L n +b]\mathbf{W}^\dagger_{\!\L}(b),\\
\overline{U}^\mu_{2,i}(z_1,z_2;b) &=& \bar{q}_i(z_1n+b)[z_1n+b,z_2n+b]gF^{\mu+}(z_2n+b)[z_2n+b,\L n +b]\mathbf{W}^\dagger_{\!\L}(b), \label{definition_SemiCompactOperators-4}
\end{eqnarray}
where $z$'s are coordinates along the light-cone direction, $b$ is the transverse separation of the fields $b^2 = -\bm{b}^2 <0$, $F_{\mu\nu}$ is the gluon-field-strength tensor, and $i$ is the spinor index of the quark-field. In Eqs.~\eqref{definition_SemiCompactOperators-1}-\eqref{definition_SemiCompactOperators-4}, the gauge link $\mathbf{W}_{\L}$ is defined along the straight path 
\begin{equation}
\begin{split}
[a,b] &= \mathcal{P}\exp\ta   -ig \int_0^1 dt (b^\mu-a^\mu) A_\mu(bt+a(1-t)) \tc , \\
\mathbf{W}_{\!\L}(b) &= \mathcal{P}\exp\ta   -ig \int_\infty^1 dt b^\mu A_\mu(bt+\L n  ) \tc.
\end{split}
\end{equation}
For all the variables with subscripts (for instance $z_i$), we will uniformly adopt the convention to reserve the subscript $2$ for the gluon variable.
The subscripts 1 and 2 of the semi-compact operators refer to the twist of these operators, where it is assumed that only the good components of the quark spinors are used to construct the operators. 

NLP TMD distributions are constructed from the semi-compact operators as
\begin{eqnarray}\label{def:PHI21}
\Phi_{\mu,21}^{[\Gamma]}(x_1,x_2,x_3,b)&=&\lim_{\L\to\pm\infty}\int_{-\infty}^\infty \frac{dz_1 dz_2}{(2\pi)^2} e^{i(x_1z_1+x_2z_2)p^+} 
\langle p,S|\overline{U}_{\mu,2}(z_1,z_2;b)\frac{\Gamma}{2}U_1(0;0)|p,S\rangle,
\\\label{def:PHI12}
\Phi_{\mu,12}^{[\Gamma]}(x_1,x_2,x_3,b)&=&\lim_{\L\to\pm\infty}\int_{-\infty}^\infty \frac{dz_1 dz_2}{(2\pi)^2} e^{i(x_1z_1+x_2z_2)p^+} 
\langle p,S|\overline{U}_{1}(z_1;b)\frac{\Gamma}{2}U_{\mu,2}(z_2,0;0)|p, S\rangle,
\end{eqnarray}
where $\Gamma\in \text{span}\{\gamma^+,\gamma^+\gamma_5,i\sigma^{\alpha+}\gamma_5\}$, with the $\alpha$ index  assumed in the transverse space. The sign of the infinity in the limit over $\L$ depends on the process under consideration: $-$ is for Drell-Yan, $+$ for SIDIS.

It turns out that, although the distributions in Eqs. \eqref{def:PHI21} and \eqref{def:PHI12} have well-defined interpretations in terms of operators, they do not have  definite complexity and T-parity (see \cite{Rodini:2022wki}).

To mitigate this problem, the $\oplus,\ominus$ distributions are introduced as follows
\[
\begin{split}
\Phi_{\mu,\oplus}^{[\Gamma]}(x_1,x_2,x_3,b) &= \frac{\Phi_{\mu,21}^{[\Gamma]}(x_1,x_2,x_3,b)+\Phi_{\mu,12}^{[\Gamma]}(-x_3,-x_2,-x_1,b)}{2},\\
\Phi_{\mu,\ominus}^{[\Gamma]}(x_1,x_2,x_3,b) &= \frac{i\Phi_{\mu,21}^{[\Gamma]}(x_1,x_2,x_3,b)-i\Phi_{\mu,12}^{[\Gamma]}(-x_3,-x_2,-x_1,b)}{2}.
\end{split}
\]
Because of the definite T-parity and complexity, these linear combinations have   a more natural interpretation as the real ($\oplus$) and imaginary ($\ominus$) part of the partonic process \cite{Rodini:2023plb} (choosing, for instance, light-cone gauge with appropriate boundary conditions for the transverse gluon field at light-cone infinity, all the Wilson lines in the TMD operators can be set to unity\footnote{It is important to stress that this is always possible for an isolated TMD distribution. However,  in physical cross sections, TMD distributions appear always in pairs and  it is not possible to choose a gauge that removes the  Wilson lines from \textit{both} TMD distributions.}).
The parametrization of the TMD distributions is conveniently  performed  for the definite T-parity combinations (where $f(x_{1,2,3})\equiv f(x_1,x_2,x_3)$) as
\begin{eqnarray}\label{def:TMDs:2:g+}
\Phi_{\bullet}^{\mu[\gamma^+]}(x_{1,2,3},b)&=&
\epsilon^{\mu\nu}s_{T\nu} M f_{\bullet T}(x_{1,2,3},b)
+ ib^\mu M^2 f^\perp_\bullet(x_{1,2,3},b)
\\\notag  &&
+i\lambda \epsilon^{\mu\nu}b_\nu M^2 f^\perp_{\bullet L}(x_{1,2,3},b)
+b^2 M^3\epsilon_T^{\mu\nu}\ta \frac{g_{T,\nu\rho}}{2}-\frac{b_\nu b_\rho}{b^2 }\tc s^{\rho}_Tf_{\bullet T}^\perp(x_{1,2,3},b),
\\\label{def:TMDs:2:g+5}
\Phi_{\bullet}^{\mu[\gamma^+\gamma^5]}(x_{1,2,3},b)&=&
s_T^\mu M g_{\bullet T}(x_{1,2,3},b)
-i\epsilon^{\mu\nu}_Tb_\nu M^2 g^\perp_\bullet(x_{1,2,3},b)
\\\notag  &&
+i\lambda b^\mu M^2 g_{\bullet L}^\perp(x_{1,2,3},b)
+b^2 M^3\ta \frac{g_T^{\mu\nu}}{2}-\frac{b^\mu b^\nu}{b^2 }\tc s_{T\nu}g_{\bullet T}^\perp(x_{1,2,3},b),
\\\notag 
\Phi_{\bullet}^{\mu[i\sigma^{\alpha+}\gamma^5]}(x_{1,2,3},b)&=&
\lambda g_{T}^{\mu\alpha} M h_{\bullet L}(x_{1,2,3},b) 
+\epsilon^{\mu\alpha}_T M h_\bullet(x_{1,2,3},b)
+ig_T^{\mu\alpha}(b\cdot s_T)M^2 h_{\bullet T}^{D\perp}(x_{1,2,3},b)
\\\notag  &&
+i(b^\mu s^\alpha_T-s_T^\mu b^\alpha)M^2h_{\bullet T}^{A\perp}(x_{1,2,3},b)
+(b^\mu \epsilon^{\alpha\beta}_Tb_\beta+\epsilon_T^{\mu\beta}b_\beta b^\alpha)M^3h_{\bullet}^\perp(x_{1,2,3},b)
\\\label{def:TMDs:2:s+} &&
+\lambda M^3 b^2 \ta \frac{g^{\mu\alpha}_T}{2}-\frac{b^\mu b^\alpha}{b^2 }\tc h_{\bullet L}^\perp(x_{1,2,3},b)
\\\notag  &&
+i(b\cdot s_T) M^2\ta \frac{g^{\mu\alpha}_T}{2}-\frac{b^\mu b^\alpha}{b^2 }\tc h_{\bullet T}^{T\perp}(x_{1,2,3},b)
\\\notag  &&
+iM^2\ta \frac{b^\mu s^\alpha_T+s_T^\mu b^\alpha}{2}-\frac{b^\mu b^\alpha}{b^2 }(b\cdot s_T)\tc h_{\bullet T}^{S\perp}(x_{1,2,3},b)
,
\end{eqnarray}
where $\bullet = \oplus,\ominus$, $M$ is the hadron mass and $s_T^\mu$ is the transverse projection of the hadron spin four-vector. In Eqs.~\eqref{def:TMDs:2:g+}-\eqref{def:TMDs:2:s+}, exactly half of the distributions are T-even and half are T-odd (see \cite{Rodini:2022wki}).

For the collinear twist-3 distributions we adopt the following parametrization \cite{Scimemi:2018mmi,Rein:2022odl}:
\begin{align}
\label{PDF_tw3_1}
& \langle p,S|g\bar q(z_1n)[z_1n,z_2n]F^{\mu+}(z_2n) [z_2n,z_3n]\gamma^+q(z_3n)|p,S\rangle 
 = 2 \epsilon^{\mu\nu}_T s_\nu (p^+)^2 M \int[dx] e^{-ip^+(x_1z_1+x_2z_2+x_3z_3)}T(x_{1,2,3}),
\\
&\langle p,S|g\bar q(z_1n)[z_1n,z_2n]F^{\mu+}(z_2n) [z_2n,z_3n]\gamma^+\gamma^5q(z_3n)|p,S\rangle 
 = 2i s_T^\mu (p^+)^2 M \int[dx] e^{-ip^+(x_1z_1+x_2z_2+x_3z_3)}\Delta T(x_{1,2,3}),
\\
&\langle p,S|g\bar q(z_1n)[z_1n,z_2n]F^{\mu+}(z_2n) [z_2n,z_3n]i\sigma^{\nu+}\gamma^5q(z_3n)|p,S\rangle 
 \notag \\
 & \qquad \qquad = 2 (p^+)^2 M \int[dx] e^{-ip^+(x_1z_1+x_2z_2+x_3z_3)}
\ta\epsilon^{\mu\nu}_T E(x_{1,2,3})+i\lambda g^{\mu\nu}_T H(x_{1,2,3})\tc
\label{PDF_tw3_3}.
\end{align}
The integral measure 
\begin{eqnarray}\label{def:[dx]}
\int [dx]=\int_{-1}^1 dx_1 d x_2 dx_3 \delta(x_1+x_2+x_3)
\end{eqnarray}
reflects momentum conservation. Different notations and conventions for these distributions exist in literature and can be found in, e.g., \cite{Kang:2008ey,Boer:1997bw,Kanazawa:2000hz,Boer:2003cm,Eguchi:2006mc}.

The twist-3 collinear distributions satisfy the following symmetry relations
\[
\label{Symm_PDF}
T(-x_{3,2,1}) = T(x_{1,2,3}), \quad \Delta T(-x_{3,2,1}) =-\Delta T(x_{1,2,3}), \quad E(-x_{3,2,1}) = E(x_{1,2,3}), \quad H(-x_{3,2,1}) =-H(x_{1,2,3}).
\]

%%%%%%%%%%%%%%%%%%%%%%%%%%
\section{Overview of the method}
\label{sec_Overview_method}
%%%%%%%%%%%%%%%%%%%%%%%%%%
The core idea of the OPE is to expand the operator of interest into a series of local operators, sort them into irreducible representations of the Lorentz group (twist) and then resum  all the operators 
with the same twist in a non-local form.
The twist decomposition can be performed using different techniques. Here we are going to employ the spinor-based formalism described in \cite{Moos:2020wvd,Braun:2008ia,Braun:2009vc}. Examples of different approaches   can be found in \cite{Scimemi:2018mmi,Ball:1998sk,Balitsky:1987bk,Geyer:1999uq,Geyer:2000ig,Belitsky:2000vx}. 
As discussed in \cite{Moos:2020wvd}, the twist decomposition for TMD distributions requires special care. 
This is related to the necessity of compactifying the TMD operator (i.e. setting $\L$ finite), enabling the expansion of the operator into a local series.  When resumming  the series in the limit $\L\to\infty$, T-odd effects arise. In the present work, this issue will not be a point of concern, as the generation of T-odd effects at tree-level
necessitates a matching onto PDFs with a different number of fields compared to the TMD operator (compare with the results in \cite{Moos:2020wvd}).

Semi-compact operators are expanded around $b_0=0$ and $z_0=\L$ :
\begin{align}
\label{U1_expansion}
U_{1,i}(z,b) &= \sum_n \frac{1}{n!}\sum_s \frac{(z-\L)^s}{s!} b^{\mu_1}\dots b^{\mu_n} \rDer{D}_{\mu_1}\dots \rDer{D}_{\mu_n} \rDer{D}_+^s q_i(\L n), \\
\label{U2_expansion}
U^\mu_{2,i,c_1}(z_2,z_1,b) &= \sum_n \frac{1}{n!}\sum_{s,t} \frac{(z_1-\L)^s}{s!}\frac{(z_2-\L)^t}{t!} b^{\mu_1}\dots b^{\mu_n} \rDer{D}_{\mu_1}\dots \rDer{D}_{\mu_n} \qa  \ta(\rDer{D}_+^t)_{ab} F_b^{\mu+}(\L n)\tc T^a_{c_1 c_2} \ta\rDer{D}_+^s q_{i,c_2}(\L n)\tc \qc,
\end{align}
where  we explicitly indicated the color index contractions ($c_{1,2}$ are in the fundamental representation, $a,b$ in adjoint representation).  Henceforth, we will suppress all color indexes. 
Additionally, we define the auxiliary variables  $w_i\equiv z_i-\L$. 
Since we are going to consider only diagonal matrix elements, we can use translational invariance to turn right derivatives into left derivatives. We have
\[
\label{Expansion_U2_U1}
\bar{U}^\mu_{2}(z_1,z_2,b) \frac{\Gamma}{2} U_{1}(z_3,0) = \sum_n \frac{b^{\mu_1}\dots b^{\mu_n}}{n!} \sum_{s,t,v} \frac{w_1^s}{s!}\frac{w_2^v}{v!}\frac{w_3^t}{t!} \qa\bar{q} D_+^s \ta F^{\mu+} D_+^v\tc D_{\mu_1}\dots D_{\mu_n} \frac{\Gamma}{2} \rDer{D}_+^t q \qc(\L n),
\]
where, if  not specified, all derivatives act on the left.
In this form, we can easily translate the expansion in the spinor form.

The underlying principle is to utilize the local isomorphism between the Lorentz group and $SL(2,\mathbb{C})$ in order to isolate,  within the series of local operators, the contributions  corresponding  to different representations of the Lorentz group.
The isomorphism is realized using $\sigma^{\mu} = \{1,\sigma^1,\sigma^2,\sigma^3\}$, where $\sigma^i$ are the Pauli matrixes and
\[
x_{\alpha \dot{\alpha}} = x_\mu \sigma^{\mu}_{\alpha \dot{\alpha}}.
\]
Dotted and undotted indexes are not equivalent since they belong to conjugate representations, i.e. $(u_\alpha)^* = \bar{u}_{\dot{\alpha}}$. 
The antisymmetric tensor is used to raise and lower indexes and  $\e_{12} = -\e_{\dot{1}\dot{2}} = 1$. Scalar products are therefore anti-symmetric: $(uv) = u^\alpha v_\alpha = -(vu)$ and $(\bar{u}\bar{v}) = \bar{u}_{\dot{\alpha}}v^{\dot{\alpha}}=-(\bar{u}\bar{v})$. For further details, we refer to \cite{Moos:2020wvd} and references therein.

For our purposes, let us simply recall that the light-cone vectors can be parametrized in the spinor representation using two spinors $\lambda$ and $\mu$, such that
\[
n_{\alpha\dot{\alpha}} = \lambda_\alpha \bar{\lambda}_{\dot{\alpha}}, \quad \bar{n}_{\alpha\dot{\alpha}} = \mu_\alpha \bar{\mu}_{\dot{\alpha}}.
\]
This implies the normalization condition $(\mu\lambda)(\bar{\lambda}\bar{\mu}) = 2$. 
These spinors are sufficient to fully parametrize any four vector:
\[
v_{\alpha\dot{\alpha}} = (v\bar{n}) \lambda_\alpha \bar{\lambda}_{\dot{\alpha}} + (vn)\mu_\alpha \bar{\mu}_{\dot{\alpha}} - v_T \mu_\alpha \bar{\lambda}_{\dot{\alpha}} - \bar{v}_T \lambda_\alpha\bar{\mu}_{\dot{\alpha}}.
\]
For the sake of brevity, we adopt the convention $v_{\alpha\dot{\alpha}} \lambda^\alpha \bar{\lambda}^{\dot{\alpha}} \equiv v_{\lambda\bar{\lambda}}$ and similar for the other contractions.
In the spinor space, since the scalar product is anti-symmetric, any symmetric tensor in the spinor indexes is necessarily traceless. Hence, to project any given Lorentz representation, it is sufficient to either symmetrize or  anti-symmetrize the spinor indexes. The lowest twist component, which is the only one of interest, is obtained when all indexes are symmetrized. To symmetrize the spinor indexes of a generic operator of the form
\[
\Gamma^{\alpha_1...\alpha_n}(\lambda,\mu)\bar{\Xi}^{\dot{\alpha}_1...\dot{\alpha}_m}(\bar{\lambda},\bar{\mu}) \mathcal{O}_{\alpha_1...\alpha_n,\dot{\alpha}_1...\dot{\alpha}_m},
\]
we first replace all  occurrences of $\mu$ with $\lambda$. This obviously symmetrizes the tensor. 
Next, we replace as many occurrences of $\lambda$ with  $\mu$ as needed  to restore the original count of $\mu$. 
If in the original tensor we have $N$ occurrences of $\mu$ and $L$ occurrences of $\lambda$, then we have $N!$ identical ways to perform the first step. For the second step we have $(L+N)(L+N-1)...(L+1) = (L+N)!/L!$ ways to select the $\lambda$'s to be replaced by $\mu$'s. Hence, the normalized symmetric projector reads
\[
S_{N} = \frac{L!}{N!(L+N)!}\ta\frac{\mu\partial}{\partial \lambda}\tc^N \ta \frac{\lambda\partial}{\partial \mu}\tc^{N},
\]
where the derivatives 
\[
\frac{\lambda\partial}{\partial \mu} \equiv \lambda^\alpha \frac{\partial}{\partial \mu^\alpha}
\]
perform the substitutions.
Combining the symmetrization in both spinor representations (with $\bar{L},\bar{M}$ the number of $\bar{\lambda},\bar{\mu}$ occurrences, respectively), we have that the leading twist projection operator reads\footnote{In \cite{Moos:2020wvd} the operator corresponding to the lowest twist was denoted as $T_2$, with reference to the fact that in the specific case studied, the lowest twist was twist-2. To prevent any potential confusion, we renamed it as $\mathbb{T}_{\text{LT}}$,  as the lower twist in our case is actually twist-3.} 
\[
\mathbb{T}_{\text{LT}} = \frac{L!}{N!(L+N)!}\frac{\bar L!}{\bar N!(\bar L+\bar N)!}\ta\frac{\mu\partial}{\partial \lambda}\tc^N \ta\frac{\bar\mu\partial}{\partial \bar\lambda}\tc^{\bar{N}} \ta \frac{\lambda\partial}{\partial \mu}\tc^{N} \ta \frac{\bar \lambda\partial}{\partial \bar\mu}\tc^{\bar N}.
\]
As a final remark, an alternative approach could have been to symmetrize the operator by substituting all occurrences of $\lambda$ with $\mu$, which would have yielded equivalent results. However, this alternative is highly impractical in the present case.

As a final step, one has to write the fields in the spinor space.  We have for the fermions and for the gluons (assuming one light-cone and one transverse index)
\[
q = \begin{pmatrix}
\psi_\alpha\\
\bar{\chi}^{\dot{\beta}}
\end{pmatrix}, \quad  F_{\alpha\dot{\alpha},+} = \frac{1}{2}F_{\alpha\dot{\alpha},\l \lb} = -\mu_\alpha\lb_{\dot{\alpha}}\frac{f_{\l\l}}{(\mu\l)}-\l_\alpha \mb_{\dot{\alpha}}\frac{\bar{f}_{\lb\lb}}{(\lb\mb)}.
\] 
Similarly, for the derivatives we have:
\[
D_+ = \frac{D_{\l\lb}}{2}, \quad b^\mu D_\mu = -\frac{1}{2}\ta b_T D_{\mu\lb} + \bar{b}_TD_{\l\mb}\tc.
\]
We can now express Eqs.~\eqref{U1_expansion} and \eqref{U2_expansion} in the spinor representation, apply $\mathbb{T}_{\text{LT}}$ and combine the results into the form of the correlator projections in Eqs.~\eqref{def:TMDs:2:g+}-\eqref{def:TMDs:2:s+}. The application of the $\mu\partial/\partial\lambda$ derivatives and the subsequent resummation of the series is more conveniently performed at the level of the parametrization for the, now collinear, matrix elements \eqref{PDF_tw3_1}-\eqref{PDF_tw3_3}. The technique and caveats of the computation can be found in \cite{Moos:2020wvd}.

In principle, one could also compute the next-to-lower twist contributions by anti-symmetrizing one pair of indexes and symmetrizing all the others. However, in the present case, this approach generates  twist-4 operators. It is well-known (see \cite{Braun:2008ia}) that starting from twist-4,  non-quasi-partonic operators must be considered. This considerably  
complicates the computation and, for all foreseeable applications of the small-$b$ matching, twist-4 collinear distributions are deemed  irrelevant.
%%%%%%%%%%%%%%%%%%%%%%%%%%
\section{Results}
\label{sec_Results}
%%%%%%%%%%%%%%%%%%%%%%%%%%
All the results are derived in the form of the following convolution
\[
\mathcal{C}[g(u)F(y_{1,2,3})]\equiv\int [dy] \int_0^1 du \ g(u) F(y_1,y_2,y_3) \delta(x_1-uy_1)\delta(x_2-uy_2) .
\]
This could be explicitly written as a sum of integrals over $u$, by virtue of the support property $x_i\in[-1,1]$ and $y_i\in[-1,1]$ of the collinear distributions.

Only a very small fraction of the NLP TMD parton distributions exhibits a non-zero tree-level matching contribution. As an example, we show the derivation of the results for $f_{12,T}$ contribution in~\ref{appendix}. Similarly, one can deduce the results for all the other contributions. The computation leads to the following results:  
\begin{align}
\label{f_plus_T}
f_{\oplus,T} &= T(x_{1,2,3}) +\sum_{k=1}^\infty \ta \frac{x_3^2b^2  M^2}{4}\tc^k    \frac{1}{\Gamma (k)\Gamma(k+2) } 
\mathcal{C}\qa (\bar{u}+u(k+u) )\ta \frac{\bar{u}}{u}\tc^{k-1}  T(y_{1,2,3}) \qc, \\
g_{\ominus,T} &= -\Delta T(x_{1,2,3}) -\sum_{k=1}^\infty \ta \frac{x_3^2b^2  M^2}{4}\tc^k    \frac{1}{\Gamma (k)\Gamma(k+2) } 
\mathcal{C}\qa (\bar{u}+u(k+u) )\ta \frac{\bar{u}}{u}\tc^{k-1}  \Delta T(y_{1,2,3}) \qc,\\
%%%%%
f_{\oplus,L}^\perp &= x_3\sum_{k=0}^\infty \ta\frac{x_3^2b^2 M^2}{4}\tc^{k}     \frac{ 1 }{ \Gamma (k+1)\Gamma(k+2)} 
   \mathcal{C}\qa u(k+u)\ta\frac{\bar{u}}{u}\tc^{k}  T(y_{1,2,3})\qc, \\
%%%%
g_{\ominus,L}^\perp  &= -x_3\sum_{k=0}^\infty \ta\frac{x_3^2b^2 M^2}{4}\tc^{k}  \frac{1 }{ \Gamma (k+1)\Gamma(k+2)} 
  \mathcal{C}\qa u(k+u) \ta\frac{\bar{u}}{u}\tc^{k}  \Delta T(y_{1,2,3})\qc, \\
%%%%%
h_{\oplus} &= E(x_{1,2,3}) + \sum_{k=1}^\infty \ta \frac{x_3^2 b^2 M^2}{4}\tc^k    \frac{1 }{\Gamma (k+1) \Gamma (k) } \mathcal{C}\qa u\ta \frac{\bar{u}}{u}\tc^{k-1} E(y_{1,2,3}) \qc, \\
%%%%%
h_{\ominus,L} &= -H(x_{1,2,3})-\sum_{k=1}^\infty \ta\frac{x_3^2b^2  M^2}{4}\tc^{k}   \frac{1}{\Gamma(k)\Gamma(k+1)}  \mathcal{C}\qa u(1+k\bar{u}\Phi(\bar{u},1,k))\ta\frac{\bar{u}}{ u}\tc^{k-1} H(y_{1,2,3})\qc, \\
%%%%%
h_{\ominus,T}^{D\perp} &= -x_3\sum_{k=0}^\infty \ta \frac{x_3^2b^2 M^2}{4}\tc^k \frac{1}{\Gamma^2(k+1)}\mathcal{C}\qa   {}_2F_1(1,1;1+k;\bar{u})  \ta\frac{\bar{u}}{u}\tc^{k} u^2 H(y_{1,2,3})\qc,
\end{align}
where $\Phi(\bar{u},1,k)$ is the Lerch function.
We can see the effect of this matching on the SIDIS structure functions. Let us define the leading order convolutions (see \cite{Rodini:2023plb})
\begin{align}
\mathcal{J}^{[R]}_n[f_\bullet D]&=x\sum_i e_i^2  \int_0^\infty \frac{b db}{2\pi} (bM)^n J_n\ta\frac{b|p_\perp|}{z}\tc
\int_{-1}^{1-x}du_2 \frac{1}{(u_2)_+}  f_{i\bullet}(-x-u_2,u_2,x,b;\mu,\zeta)D_i(z,b;\mu,\bar \zeta),
\\
\mathcal{J}^{[I]}_n[f_\bullet D]&=x\pi\sum_i e_i^2  \int_0^\infty \frac{b db}{2\pi} (bM)^n J_n\ta\frac{b|p_\perp|}{z}\tc f_{i\bullet}(-x ,0,x,b;\mu,\zeta)D_i(z,b;\mu,\bar \zeta),
\end{align}
where the plus distribution is
\begin{eqnarray}
\int du (f(u))_+g(u)=
\int du f(u)(g(u)-g(0)),
\end{eqnarray}
and where $f_{i\bullet}$ is a generic NLP TMD PDF for the proton and $D_i$ is a generic LP TMD fragmentation function for a hadron of mass $m_h$.

\begin{table}[!ht]
\centering
\begin{tabular}{c|c}
$F^{\cos\phi}_{UU} $ 
   &  $\frac{4m_h}{Q} \mathcal{J}^{[I]}_1[   h_\oplus H_1^\perp]$ \\  
$F^{\sin\phi}_{LU}  $        &  $\frac{4m_h}{Q}  \mathcal{J}^{[R]}_1[ h_\oplus H_1^\perp]$ \\
\hline
%%%%%%%%%%%%%%%%%%%%%%%%%%%%%%%%%%%%%%%%%%%%%%%%%%%%%
$F^{\sin\phi}_{UL} = $    &  $-\frac{4m_h}{Q}  \mathcal{J}^{[R]}_1[h_{\ominus L}H_1^\perp] + \frac{2M}{Q}  \mathcal{J}^{[I]}_1[ D_1f_{\oplus L}^\perp ]$ \\  
\hline
$F^{\cos\phi}_{LL}  $        &  $-\frac{2M}{Q} \mathcal{J}^{[R]}_1[ D_1(f_{\oplus L}^\perp+g_{\ominus L}^\perp) ]-\frac{4m_h}{Q} \mathcal{J}^{[I]}_1[ h_{\ominus L}H_1^\perp]$ \\
\hline
$F^{\sin\phi_S}_{UT}  $      &   $\frac{2M}{Q}\mathcal{J}^{[R]}_0\qa   Mm_hb^2   H_1^\perp h_{\ominus T}^{D\perp} \qc -\frac{2M}{Q}\mathcal{J}^{[I]}_0\qa D_1f_{\oplus T} \qc$ \\
%%%%%%%%%%%%%%%%%%%%%%%%%%%%%%%%%%%%%%%%%%%
$F^{\sin2\phi-\phi_S}_{UT} $       &  $-\frac{2m_h}{Q}\mathcal{J}^{[R]}_2[H_1^\perp h_{\ominus T}^{D\perp}]$\\
\hline
$F^{\cos\phi_S}_{LT} $  &  $\frac{2M}{Q}\mathcal{J}^{[R]}_0\qa D_1(f_{\oplus T}+g_{\ominus T}) \qc $\\
\end{tabular}
\caption{Summary of the contributions of TMD PDFs to the NLP SIDIS structure functions in the small-$b$ tree level matching (see \cite{Rodini:2023plb,Bacchetta:2006tn}). $H_1^\perp$ and $D_1$ are the LP TMD fragmentation functions and $m_h$ is the mass of the fragmenting hadron.}
\label{table_SIDIS}
\end{table}

In Tab.\ref{table_SIDIS} we collect the contributions that have non-vanishing leading-order small-$b$ matching from the  NLP TMD PDFs to the SIDIS structure functions.
For all the structure functions at next-to-leading order we refer to \cite{Rodini:2023plb}. The symmetry relations for the twist-3 PDFs \eqref{Symm_PDF}  constrain the set of distributions that can participate in  $\mathcal{J}^{[I]}$. 

In the case of an unpolarized proton ($F_{UU}$ and $F_{LU}$), only the $h_\oplus$ distribution contributes. For the totally unpolarized case $F_{UU}^{\cos \phi}$ (also known as the Cahn effect), in the small-$b$ limit, the structure function has a very similar expression to the $F_{UU}^{\cos 2\phi}$ structure function (which involves solely LP TMDs). This is because the TMD $h_\oplus$ is written as the Qiu-Stermann-like $E(-x,0,x)$, which is the same as for the twist-2 Boer-Mulders TMD. This provides an indication of the potential magnitude of non-perturbative contributions to NLP structure functions. 
For the longitudinally polarized lepton case,   it is worth noting  that $F_{LU}^{\sin\phi}$ is the only NLP structure function that does not exhibit contributions from the kinematic power corrections part (compare the full expressions in \cite{Rodini:2023plb}). Therefore  this structure function serves as a clear channel for observing the effects of the NLP TMDs.

In the case of longitudinally polarized proton ($F_{UL}$ and $F_{LL}$), we start to see the interplay between different sources of genuine NLP TMD PDFs. In this case, each structure function contains contributions from both $\mathcal{J}_1^{[R]}$ and $\mathcal{J}_1^{[R]}$. Extracting TMDs from the longitudinally polarized proton structure functions will probably be an extremely complex task, due to the necessity of disentangling the different non-perturbative effects.

The case of transversely polarized proton ($F^{\sin\phi_S}_{UT},F^{\sin2\phi-\phi_S}_{UT},F_{LT}^{\cos\phi_S}$) appears to be simpler.  However, the $\sin\phi_S$ structure function presents similar challenges as in the case of longitudinally polarized proton, since it is a combination of different NLP TMD PDFs already at the tree-level. $F^{\sin2\phi-\phi_S}_{UT}$ is better, since it provides access, via the matching, to the twist-3 PDF $H$.
This is in contrast to the matching for the LP TMD parton distributions, where $H$ appears solely in the worm-gears distributions through a rather complicated expression  (compare with \cite{Rein:2022odl}).

As a final remark, we stress that some of the structure functions that do not have a tree-level matching will receive matching at the one-loop level. Some of these structure functions will receive a `backward' matching, where the NLP TMD distributions are obtained in terms of the twist-2 (quark-quark or gluon-gluon) PDFs \cite{Rodini:2022wki}.

%%%%%%%%%%%%%%%%%%%%%%%%%%
\section{Summary}
\label{s:summary}
%%%%%%%%%%%%%%%%%%%%%%%%%%
In this work, we have derived the small-$b$ matching relations for the twist-(1,2) and (2,1) TMD parton distributions. The final results include the complete series of target mass corrections, which are found to be relatively small when $b$ is of the order of the hadron mass $M$.  We observed that, at tree-level, only a limited number of distributions have matching to twist-3 collinear distributions. As expected, at leading twist in the matching, all the T-odd effects vanish. To generate T-odd effects, one should go beyond the leading-twist approximation. In the present case, this entails  to go to twist-4 collinear distributions, which are mostly unknown and are unlikely to have a significant impact on phenomenological extractions.
From the results of the small-$b$ matching, we observed a significant simplification in the SIDIS structure functions. This simplification has clear implications  for the feasibility of phenomenological extractions of NLP TMDs from SIDIS data. With the upcoming experimental programs, we anticipate that power corrections will play a significant role in increasing the precision of the phenomenological extractions. 

%%%%%%%%%%%%%%%%%%%%%%%%%%
\section*{Acknowledgement}
The work of B.P. is part of a project that has received funding from the European Union’s Horizon 2020 research and innovation programme under grant agreement STRONG – 2020 - No~824093.

\appendix
\section{An example}\label{appendix}
Let us take a specific example to show some of the calculation details. To do so, we start by taking a specific Lorentz projection of Eq.~\eqref{Expansion_U2_U1}, namely the one that isolates the $f_{\l\l}$ component, and we choose $\Gamma = \gamma^+$. This, using the parametrization in Eq.~\eqref{def:TMDs:2:g+}, gives 
\[
\begin{split}
& -iS_{\l\mb}M p_+^2\int [dw] e^{-ip_+\sum_{i=1}^3 w_iz_i}f_{21,T}(w_{1,2,3})\\
&\overset{\mathbb{T}_{\text{LT}}}{\longrightarrow} \sum_{n=0}^\infty \frac{1}{\Gamma(n+1)} \sum_{k=0}^n \frac{\Gamma(n+1)}{\Gamma(k+1)\Gamma(n-k+1)}\ta \frac{b_T}{2}\tc^k \ta\frac{\bar{b}_T}{2}\tc^{n-k}  \\& \times\frac{\Gamma (k+m+2) \Gamma (n-k+m+4)}{\Gamma (k+1) \Gamma (n-k+1) \Gamma (m+n+2) \Gamma (m+n+4)} \Gamma(k+1)\Gamma(n-k+1)\\&
\times\sum_{s,t,v=0}^\infty \frac{(z_1-\L)^s (z_2-\L)^v (z_3-\L)^t}{s!v!t! 2^s 2^t 2^v} \ta \frac{\m\partial}{\partial \l}\tc^k \ta\frac{\mb\partial}{\partial \lb}\tc^{n-k} \bar{q}D_{\l\lb}^s \ta \frac{f_{\l\l}}{\sqrt{2}}D_{\l\lb}^v\tc  \rDer{D}_{\l\lb}^{n+t} \gamma^+ q \\
%%%%%%%%%%%%%%%%%%%%%%%%%
& \overset{\braket{p,S|.|p,S}}{\longrightarrow} \sum_{n=0}^\infty  \sum_{k=0}^n \ta \frac{b_T}{2}\tc^k \ta\frac{\bar{b}_T}{2}\tc^{n-k}   \frac{\Gamma (k+m+2) \Gamma (n-k+m+4)}{\Gamma (m+n+2) \Gamma (m+n+4)\Gamma(k+1)\Gamma(n-k+1)} \frac{M}{4}S_{\l\mb}\\&
\times\sum_{s,t,v=0}^\infty \frac{(z_1-\L)^s (z_2-\L)^v (z_3-\L)^t}{s!v!t! 2^s 2^t 2^v} \ta \frac{\m\partial}{\partial \l}\tc^k \ta\frac{\mb\partial}{\partial \lb}\tc^{n-k} p_{\l\lb}^{s+t+v+n+2}\int [dw]  (-iT(w_{1,2,3})) (-iw_1)^s (-iw_2)^v (-iw_3)^{n+t} ,
\end{split}
\]
where the symbols above the arrows indicate which operation is performed.
Now, the cases for even-$n$ and odd-$n$ must be treated separately. The former contributes to $f_{\oplus,T}$, while the latter contributes to $f_{\oplus,L}^\perp$. Considering only even values of $n$, it forces $n=2k$ and no derivative acting on the spin vector, since the proton momentum has no transverse components $p_{\l\mb}=p_{\m\lb}=0$. 
The action of the derivatives on the proton momentum reads:
\[
\ta \frac{\m\partial}{\partial \l}\tc^k \ta\frac{\mb\partial}{\partial \lb}\tc^{n-k} p_{\l\lb}^{s+t+v+n+2} = p_{\l\lb}^{s+t+v+n+2-k} p_{\m\mb}^{k} \frac{\Gamma(s+t+v+n+3)}{\Gamma(k+1)\Gamma(s+t+v+n-k+3)}\Gamma(k+1)\Gamma(n-k+1).
\]
Using the on-shellness relation for the proton, i.e. $
p_{\m\mb} = M^2/p_+
$,
one obtains
\[
p_{\l\lb}^{s+t+v+n+2-k} p_{\m\mb}^{k} = 4(2M^2)^k 2^s 2^t 2^v p_+^{s+t+v+2} ,
\]
leading to
\[
\begin{split}
& \int [dw] e^{-ip_+\sum_{i=1}^3 w_iz_i}f_{21,T}(w_{1,2,3})\\
& = \sum_{k=0}^\infty \ta \frac{2b_T\bar{b}_T M^2}{4}\tc^k    \frac{\Gamma (k+s+t+v+2) \Gamma (s+t+v+k+4)}{\Gamma (s+t+v+2k+2) \Gamma (s+t+v+2k+4)\Gamma(k+1)} \frac{\Gamma(s+t+v+2k+3)}{\Gamma(s+t+v+k+3)} \\&
\times\sum_{s,t,v=0}^\infty \frac{(z_1-\L)^s (z_2-\L)^v (z_3-\L)^t}{s!v!t! } \int [dw] T(w_{1,2,3}) (-iw_1p_+)^s (-iw_2p_+)^v (-iw_3p_+)^{t} (-iw_3)^{2k}  \\ 
& =\int [dw] e^{-ip_+\sum_{i=1}^3 w_iz_i} T(w_{1,2,3}) \\
& + \sum_{k=1}^\infty \ta \frac{-b^2 M^2}{4}\tc^k    \ta\frac{\beta (k+s+t+v+4,k)}{ \Gamma (k) \Gamma (k+1)}+\frac{ \beta (k+s+t+v+4,k)}{(k+s+t+v+2) \Gamma^2 (k) }\tc \\&
\times\sum_{s,t,v=0}^\infty \frac{(z_1-\L)^s (z_2-\L)^v (z_3-\L)^t}{s!v!t! } \int [dw]  T(w_{1,2,3}) (-iw_1p_+)^s (-iw_2p_+)^v (-iw_3p_+)^{t} (-iw_3)^{2k}. 
\end{split}
\]
In the result, we have isolated the $k=0$ term, which is already in its final form and will not be further considered in the rest of the calculation. 
Using the integral representation of the $\beta$ function, one obtains
\[
\begin{split}
& \int [dw] e^{-ip_+\sum_{i=1}^3 w_iz_i}f_{21,T}(w_{1,2,3})\\
& = \sum_{k=1}^\infty \ta \frac{b^2 M^2}{4}\tc^k   \sum_{s,t,v=0}^\infty \frac{(z_1-\L)^s (z_2-\L)^v (z_3-\L)^t}{s!v!t! }  \ta\frac{1}{ \Gamma (k) \Gamma (k+1)}+\frac{1}{(k+s+t+v+2) \Gamma^2 (k) }\tc \\&
\times \int [dw] \int du \ta \frac{\bar{u}}{u}\tc^{k-1}   u^2T(w_{1,2,3})(-iuw_1p_+)^s (-iuw_2p_+)^v (-iuw_3p_+)^{t} (-iuw_3)^{2k}  .
\end{split}
\]
Then, the two combinations of $\Gamma$ functions should be treated differently, because the first one is independent of  $s,t,v$, whereas the second one does depend on these indexes and therefore entangles the series.
For the first contribution one simply has
\[
\label{first_piece}
\begin{split}
& \sum_{k=1}^\infty \ta \frac{ b^2  M^2}{4}\tc^k    \frac{1}{ \Gamma (k) \Gamma (k+1)}\int [dw] \int_0^1 du \ta u^2w_3^2\tc e^{-ip_+ u \sum_{i=1}^3 w_iz_i}\ta \frac{\bar{u}}{u}\tc^{k-1} u^2 T(w_{1,2,3}).
\end{split}
\]

For the second term, first we use the integral representation
\[
\frac{1}{A} = \int_0^1 \frac{d\alpha}{\alpha^2} \alpha^{A+1}
\]
which gives
\[
\begin{split}
& \sum_{k=1}^\infty \ta \frac{-b^2 M^2}{4}\tc^k    \frac{1}{\Gamma^2 (k) } 
\sum_{s,t,v=0}^\infty \frac{y_1^s y_2^v y_3^t}{s!v!t! } \int [dw] \int_0^1 du \int_0^1 \frac{d\alpha}{\alpha^2} \\&
\times \ta \frac{\bar{u}}{\alpha u}\tc^{k-1} 
    \alpha^2u^2T(w_{1,2,3})(-i\alpha uw_1p_+)^s (-i\alpha uw_2p_+)^v (-i\alpha uw_3p_+)^{t} (-i\alpha uw_3)^{2k} . 
\end{split}
\]
Then, we decouple the dependence on  $u$ and $\alpha$ in the integrated function using
\[
\begin{split}
&\int_0^1 du \int_0^1 \frac{d\alpha}{\alpha^2} f(u\alpha,\bar{u}) = \int_0^1 du \int_u^1 d\alpha \frac{f\ta u, 1-\alpha\tc}{u^2} \alpha .
\end{split}
\]
This leads to
\[
\begin{split}
& \sum_{k=1}^\infty \ta \frac{b^2  M^2}{4}\tc^k    \frac{1}{\Gamma^2 (k) } 
 \int [dw] \int_0^1 du \ \ta u^2w_3^2\tc^k e^{-iup_+\sum_{i=1}^3w_iz_i}  T(w_{1,2,3}) \int_u^1 d\alpha\  \alpha   
\ta \frac{1-\alpha}{u}\tc^{k-1}  ,
\end{split}
\]
where the $\alpha$ integral can be taken, leading to
\[
\label{second_piece}
\begin{split}
&  \sum_{k=1}^\infty \ta \frac{b^2  M^2}{4}\tc^k    \frac{1}{\Gamma (k)\Gamma(k+1) } 
 \int [dw] \int_0^1 du \ \ta u^2w_3^2\tc^k e^{-iup_+\sum_{i=1}^3w_iz_i} \frac{\bar{u} (1+k u)}{1+k}   
\ta \frac{\bar{u}}{u}\tc^{k-1}  T(w_{1,2,3}) .
\end{split}
\]
Taking now the sum of Eq.\eqref{first_piece} and Eq.\eqref{second_piece}, and restoring the term from $k=0$, we obtain
\[
\begin{split}
& \int [dw] e^{-ip_+\sum_{i=1}^3 w_iz_i}f_{21,T}(w_{1,2,3}) = \int [dw] e^{-ip_+\sum_{i=1}^3 w_iz_i}T(w_{1,2,3})\\
& + \sum_{k=1}^\infty \ta \frac{b^2  M^2}{4}\tc^k    \frac{1}{\Gamma (k)\Gamma(k+2) } 
 \int [dw] \int_0^1 du \ \ta u^2w_3^2\tc^k e^{-iup_+\sum_{i=1}^3 w_iz_i}\ta \bar{u}+u(k+u)\tc   
\ta \frac{\bar{u}}{u}\tc^{k-1}   T(w_{1,2,3}).
\end{split}
\]
Repeating the same calculation for $f_{12,T}$ and taking the Fourier transformation from $z_{1,2,3}$ to $x_{1,2,3}$ one obtains Eq.\eqref{f_plus_T}.
For the $\Gamma=\gamma^+\gamma_5$ and $\Gamma=i\sigma^{\alpha+}\gamma_5$ the computation follows the same steps. For the latter, the $\alpha$ integral leads to the hypergeometric and Lerch functions that appears in the result.

%\green{The computation for the longitudinally polarized case is very close to this example. For the transversely polarized case, it is sufficient to note that only the contractions with $\bar e^\mu e^\nu$ and $e^\mu \bar e^\nu$ survive in the matrix elements \eqref{PDF_tw3_1}-\eqref{PDF_tw3_3}; after that, the computation is similar to this example.}

%\bibliography{biblio}

%\vskip3pt

\end{document}